\input harvmac
\noblackbox

\def\bfone{\relax{\rm 1\kern-.35em 1}}
\def\inbar{\vrule height1.5ex width.4pt depth0pt}

\def\IC{\relax\,\hbox{$\inbar\kern-.3em{\rm C}$}}
\def\ID{\relax{\rm I\kern-.18em D}}
\def\IF{\relax{\rm I\kern-.18em F}}
\def\IH{\relax{\rm I\kern-.18em H}}
\def\II{\relax{\rm I\kern-.17em I}}
\def\IN{\relax{\rm I\kern-.18em N}}
\def\IP{\relax{\rm I\kern-.18em P}}
\def\IQ{\relax\,\hbox{$\inbar\kern-.3em{\rm Q}$}}
\def\us#1{\underline{#1}}
\def\IR{\relax{\rm I\kern-.18em R}}
\font\cmss=cmss10 \font\cmsss=cmss10 at 7pt
\def\ZZ{\relax\ifmmode\mathchoice
{\hbox{\cmss Z\kern-.4em Z}}{\hbox{\cmss Z\kern-.4em Z}}
{\lower.9pt\hbox{\cmsss Z\kern-.4em Z}}
{\lower1.2pt\hbox{\cmsss Z\kern-.4em Z}}\else{\cmss Z\kern-.4em
Z}\fi}
\def\a{\alpha} \def\b{\beta}

 \def\s{\sigma}\def
\om{\omega}\def\ttau{\tilde\tau}

\def\cF{{\cal F}}

\def\nup#1({Nucl.\ Phys.\ $\us {B#1}$\ (}
\def\atmp#1({Adv.\ Theo. \& Math. Phys.\ $\us {#1}$\ (}
\def\plt#1({Phys.\ Lett.\ $\us  {B#1}$\ (}
\def\cmp#1({Comm.\ Math.\ Phys.\ $\us  {#1}$\ (}
\def\prp#1({Phys.\ Rep.\ $\us  {#1}$\ (}
\def\prl#1({Phys.\ Rev.\ Lett.\ $\us  {#1}$\ (}
\def\prv#1({Phys.\ Rev.\ $\us  {#1}$\ (}
\def\mpl#1({Mod.\ Phys.\ Let.\ $\us  {A#1}$\ (}
\def\ijmp#1({Int.\ J.\ Mod.\ Phys.\ $\us{A#1}$\ (}
\def\jag#1({Jour.\ Alg.\ Geom.\ $\us {#1}$\ (}
\def\tit#1|{{\it #1},\ }

\def\Coe#1.#2.{{#1\over #2}}

\def\coe#1.#2.{\relax{\textstyle {#1 \over #2}}\displaystyle}
\def\half{{1 \over 2}}

\def\del{\partial}

\def\br{\hfill\break}

\def\tE{\widetilde E}
\def\ctF{\widetilde\cF}
\def\bE{\overline{E}}

\def\dE{\del_{E_2}}
\def\dpi{\del_{\phi_i}}
\def\tq{\tilde q}

%
%
%
\lref\KV{S.~ Kachru and C.~ Vafa, \nup{450} (1995) 69}
\lref\KKLMV{S.~ Kachru, A.~Klemm, W.~Lerche, P.~Mayr and
C.~Vafa, \nup{459} (1996) 537}
\lref\KLMVW{A.\ Klemm, W.\ Lerche, P.\ Mayr, C.\ Vafa and
N.P.\ Warner,\nup{477} (1996) 746, hep-th/9604034.}
\lref\MNW{J.A.~Minahan, D.~Nemeschansky and
N.P.~Warner, {\it Investigating the BPS Spectrum of Non-Critical
$E_n$ Strings,}  {\it to appear in Nucl. Phys. B},
USC-97/006, NSF-ITP-97-055, hep-th/9705237.}
\lref\LMW{W.~Lerche, P.~Mayr and
N.P.~Warner, {\it Non-Critical Strings, Del Pezzo Singularities
and Seiberg-Witten Curves,} \nup{499} (1997) 125, hep-th/9612085.}
\lref\KMV{A.\ Klemm, P.\ Mayr and C.\ Vafa,
{\it BPS states of exceptional non-critical strings,}
CERN-TH-96-184, hep-th/9607139.}
\lref\SW{N.\ Seiberg and E.\ Witten, \nup{426} (1994) 19,
hep-th/9407087; \nup{431} (1994) 484, hep-th/9408099.}
\lref\EnNCS{O.\ Ganor and A.\ Hanany, \nup{474} (1996) 122,
hep-th/9602120; \br
N.\ Seiberg and E.\ Witten, \nup{471} (1996) 121,
hep-th/9603003; \br
M.~Duff, H.~Lu and C.N.~Pope, \plt{378} (1996) 101,
hep-th/9603037; \br
M.R.~Douglas, S.~Katz, C.~Vafa, {\it
Small Instantons, del Pezzo Surfaces and Type I' theory,}
hep-th/9609071; \br
E.~Witten, \mpl{11} (1996) 2649, hep-th/9609159.}
\lref\MandFWit{E.~Witten, \nup{471} (1996) 195,
hep-th/9603150.}
\lref\Ganor{O.~Ganor, \nup{479} (1996) 197, hep-th/9607020;
\nup{488} (1997) 223, hep-th/9608109; {\it  Compactification of
Tensionless String Theories,}  PUPT-1634, hep-th/9607092.}
\lref\GMS{O.\ Ganor, D.\ Morrison and N.\ Seiberg,
\nup{487} (1997) 93,  hep-th/9610251.}
\lref\DMCV{D.R.~Morrison and C.~Vafa, \nup{473} (1996) 74,
hep-th/9602114;   \nup{476} (1996) 437, hep-th/9603161.}
\lref\EMa{E.\ Martinec, \plt{367}, (1996) 91, hep-th/9510294}
\lref\HIAM{H.\ Itoyama and A.\ Morozov, \nup{477} (1996) 855,
hep-th/9511126; \nup{491} (1997) 529, hep-th/9512161.}
\lref\DKMmd{N. Dorey, V. V. Khoze and M. P. Mattis,
{\it On Mass-Deformed $N=4$ Supersymmetric Yang-Mills Theory},
\plt{396} (1997) 141, hep-th/9612231}
\lref\MNWII{J.A.~Minahan, D.~Nemeschansky and
N.P.~Warner, {\it Partition Functions for BPS
States of the Non-Critical $E_8$ String,} \atmp{1} (1997) 169,
hep-th/9707149.}
\lref\DP{E. D'Hoker and D.H.~Phong, {\it Calogero-Moser Systems in
$SU(N)$ Seiberg-Witten Theory},  hep-th/9709053.}
\lref\Matone{M.~Matone, {\it Instantons and recursion relations in
$N=2$ Susy
gauge theory}, \plt{357} (1995) 342, hep-th/9506102.}
\lref\RDEW{R.~Donagi and E.~Witten, {\it Supersymmetric Yang-Mills
Systems
And Integrable Systems} \nup{460 } (1996) 299, hep-th/9510101.}
\lref\JMDN{J.A.~Minahan and D.~Nemeschansky, \nup{464}(1996) 3,
hep-th/9507032; \nup{468} (1996) 72, hep-th/9601059.}
\lref\Seib{N.~Seiberg, {\it Five Dimensional SUSY Field Theories,
Non-trivial  Fixed Points and String Dynamics}, \plt{388} (1996)
753, hep-th/9608111.}
\lref\BCOV{M. Bershadsky, S. Cecotti, H. Ooguri and  C Vafa, 
\nup{405} (1993) 279, hep-th/9302103; \cmp{165} (1994) 311, hep-th/9309140.}
\lref\BSRSW{A. Brandhuber, S. Stieberger, \nup{488} (1997) 199, hep-th/9610053;
J. Schulze, N. Warner, \nup{498} (1997) 101, hep-th/9702012;
J. Rabin, {\it Geodesics and BPS States in $N=2$ Supersymmetric QCD},
hep-th/9703145.}
%

%
\Title{\vbox{
\hbox{USC-97/016}
\hbox{\tt hep-th/9710146}
}}{\vbox{\centerline{\hbox{Instanton Expansions for Mass Deformed}}
\vskip 8 pt
\centerline{ \hbox{$N=4$ Super Yang-Mills Theories}}}}
\centerline{J.~A.~Minahan, D.~Nemeschansky 
and N.~P.~Warner}
\bigskip
\centerline{{\it Physics Department, U.S.C.}}
\centerline{{\it University Park, Los Angeles, CA 90089}}
\bigskip

\vskip .3in

\vskip .3in

We derive modular anomaly equations from the Seiberg-Witten-Donagi
curves for softly broken $N=4$ $SU(n)$ gauge theories. {}From these
equations we can derive recursion relations for the pre-potential in
powers of $m^2$, where $m$ is the mass of the adjoint hypermultiplet.
Given the perturbative contribution of the pre-potential and the
presence of ``gaps'' we can easily generate the $m^2$ expansion in
terms of polynomials of Eisenstein series, at least for relatively
low rank groups. This enables us to  determine efficiently  the
instanton expansion up to fairly high order for these gauge groups,
{\it e. g.} eighth order for $SU(3)$.  We find that after taking a derivative,
the instanton expansion of the pre-potential has integer coefficients. 
We also postulate the form of the modular anomaly equations, 
the recursion relations and the form of the instanton expansions for the
$SO(2n)$ and $E_n$ gauge groups, even though the corresponding
Seiberg-Witten-Donagi curves are unknown at this time.

\Date{\sl {October, 1997}}

%
\parskip=4pt plus 15pt minus 1pt
\baselineskip=15pt plus 2pt minus 1pt
%

\newsec{Introduction}

One of the more remarkable consequences of string duality is that it
sometimes has highly non-trivial implications for field theory limits
of the string, and it has thus led to new insights for some purely
field theoretic questions. An example of this is the use of
heterotic/type II duality to re-derive \refs{\KV\KKLMV{--}\KLMVW} the
effective actions of Seiberg and Witten for $N=2$ supersymmetric
Yang-Mills theories \SW. This re-derivation showed rather explicitly
how the Yang-Mills theory could be viewed as a compactification of
the $(0,2)$ non-critical string in six dimensions, where the
compactifying manifold is the Riemann surface of \SW, and in which (a
particular form of) the Seiberg-Witten differential represents the
local string tension on the Riemann surface. With this insight, the
study of the stable states in the BPS spectrum of the {\it field
theory} is reduced to the study of classical geodesics on this
Riemann surface\refs{\KLMVW,\BSRSW}.

Toroidal compactifications of the $E_8$, $(0,1)$ supersymmetric
non-critical strings in six dimensions have also been extensively
studied \refs{\EnNCS\MandFWit\Ganor\KMV\GMS\LMW{--}\MNW}.
In \refs{\LMW} it was shown how such string
compactifications could also be characterized by a
generalized Seiberg-Witten differential on the compactification
torus.  The slightly unusual feature was that the analogue
of the instanton expansion for the four dimensional effective
action actually counted electric BPS excitations of the
non-critical string.  This could be partially understood via the
compactification of the six dimensional theory on a circle to
five dimensions, in which four dimensional instantons have the
interpretation as five-dimensional BPS solitons \refs{\Seib}.

In \refs{\MNW} the effective action and differential were extensively
used to generate characters of the $E_n$ non-critical strings
compactified on a torus, but with the string winding around only one
of the circles. There were two natural moduli in these characters:
the string tension, which indexed the string winding number, and the
complex structure, $\tau$, of the torus, which indexed the momentum
states of the string for fixed winding number. In \refs{\MNWII}
it was shown that the partition
function, $G_n(\tau)$, of all the string momentum states of
winding number $n$ is an almost modular form in $\tau$.  That
is, $G_n$ can be written as a polynomial in the Eisenstein
functions $E_2$, $E_4$ and $E_6$.  Moreover, a recursion
relation for $G_n$ was derived, and this combined with
requirement that certain momentum states be absent, completely
determined the $G_n$, and in principle enables the computation
of the $G_n$ to arbitrary order.  These results were, of course,
consistent with the numbers generated in \KMV\ by counting
rational curves in a del Pezzo surface.

Given the similarity of the effective actions for the
non-critical strings and the effective actions of the more
usual quantum field theories, it is natural to ask whether the
instanton expansion of quantum field theories exhibits a
similar structure.  Specifically, the softly broken
$N=4$ supersymmetric gauge theories have an $SL(2,\ZZ)$
invariance when the hypermultiplet mass, $m$, is zero (and
$N=4$ supersymmetry is restored).  At this point there
are no instanton corrections to the effective action.
Turning on the mass introduces non-trivial corrections to the
effective action, but it also breaks the $S$-duality of the theory.
If one now expands the effective action in ${m \over M_j}$,
where the $M_j$ are the masses of the charged vector bosons,
one gets a series whose coefficients, $G_\alpha$, are functions
of the original $N=4$ coupling parameter, $\tau$.  In this
paper we show that these coefficient functions are, once again
almost modular forms (they be written as polynomials in the
Eisenstein functions $E_2$, $E_4$ and $E_6$).  We also show
that these coefficient functions satisfy a recursion relation.
Once again this recursion relation can be used to completely
determine the instanton expansion given two other physical pieces
of input: (i) The (known) perturbative correction to the effective
action, and (ii)  the requirement that the $m \to \infty$ limit
of the instanton series is finite.

Apart from being of theoretical interest in precisely
cataloguing the breaking of $S$-duality, the foregoing proves
of major computational significance in that it enables one
to determine the multi-instanton corrections to the effective
action rather efficiently.  We used it to determine the
$SU(2)$ effective action to instanton order 24, and the $SU(3)$
effective action to instanton order eight.
We also use it to conjecture the forms
of the instanton expansions for the softly broken $N=4$
theories with $SO(2n)$ and $E_n$ gauge groups, in spite of the
fact that the corresponding Seiberg-Witten actions are unknown.

Since gauge theories and non-critical strings are intimately
related, one would like to find a stringy interpretation of
our results, and the recursion relation in particular.  This,
as yet eludes us, but our results suggest that one should be able
to relate the instanton expansion to a topological field theory,
and a topological string theory in particular.  First, we find
if we take a $\tau$ derivative on the pre-potential, and
expand in ${m \over M_j}$ and $q = e^{2 \pi i \tau}$, then
all the coefficients are {\it integers} with uniform signs for
each monomial in ${m \over M_j}$.   This of course suggests
a topological amplitude, but whether it is a topological
Yang-Mills theory or a topological string, or both is unclear
from this observation.  Rather more ``stringy'' behaviour is
suggested by one of our results for the $SU(2)$ theory,
which will presumably generalize to other
gauge groups.  If one computes the pre-potential in the limit
as $m \to M_W$, where $M_W$ is the mass of the $W$-boson, and
subtracts the divergent perturbative piece, one finds
that the instanton series for the pre-potential is very
reminiscent of string threshhold corrections\BCOV:
\eqn\threshhold{\del_{M_W}^2 {\cal F}_{instanton} ~=~
{2 \pi i \over 24}~ \big( \log\big(\eta(2 \tau) \big) ~-~
\log\big(\eta(\tau) \big)   \big)\ .}
Thus we feel that our results will ultimately provide insight
into the formulation of Yang-Mills theories in terms of
non-critical string theories.

In section 2 we derive our recursion relation for the $SU(2)$
gauge theory, and then use it and the methods outlined above
to derive the instanton expansion.  This section is also a
model for the subsequent sections: Section 3 generalizes the
results to $SU(n)$, while section 4 discusses other simply laced
groups.  Section 5 contains some brief concluding remarks.

\newsec{Instanton expansions in the $SU(2)$ theory}

\subsec{Derivation of the Recursion Relation}

The result presented here is actually a special case of the
recursion relation derived in the next section.  However
we present a derivation here for several reasons:
(i) to make this section  self-contained,  (ii)
to provide a more explicit guide to the computation in section 3,
and (iii) to highlight the similarities with the derivation of
the recursion relation for the non-critical string \refs{\MNWII}.

The mass deformed $SU(2)$ curve is given by
\eqn\curveSUII{
y^2=\prod_{i=1}^3\left(x-e_i u- {1\over4}{e_i}^2 m^2\right),
}
where $e_i$ are the combinations of Jacobi theta functions given in
\SW.
The $SU(2)$ invariant $u$ is related to $\tr\phi^2$
by the relation\DKMmd
\eqn\ueq{
u ~=~ \tr(\phi^2) ~+~ m^2\sum_{n=0}\a_n e^{2\pi i\tau n}
}
It is not necessary to know the $\a_n$ for what follows.

The quantities $e_i$ can be rewritten in terms of Eisenstein series,
with
\eqn\Eisen{
E_4={3\over2}\sum_i {e_i}^2\qquad E_6=-{9\over2}\sum_{i\ne
j}e_i{e_j}^2
}
If we define a new variable $\chi=m^2/(12u)$, then after a rescaling
of
$x$ and $y$ the mass deformed curve
can be rewritten as
\eqn\curverw{\eqalign{
y^2=4x^3-{4\b^4u^2\over 3}&(E_4+2\chi E_6+\chi^2{E_4}^2)x\cr
&-
{8\b^6u^3\over27}(E_6+3{E_4}^2\chi+
3E_4E_6\chi^2+(2{E_6}^2-{E_4}^3)\chi^3),}
}
where the scale $\b$ will be fixed below.

We now compare the curve in \curverw
to the standard elliptic curve
\eqn\ellcurve{
y^2=4x^3-{g_2(\ttau)\over\om^4}x-{g_3(\ttau)\over\om^6}
}
where $\om$ is one of the periods and $\ttau$ is the modulus for
the curve, which depends on $\tau$ and $\chi$.
 Using $g_2(\ttau)={4\over3}\pi^4\tE_4$ and
$g_3(\ttau)={8\over27}\pi^6\tE_6$ and comparing \curverw\ with
\ellcurve,
we see that
\eqn\omeq{
\om={\pi\over\b u^{1/2}}
\left({\tE_4\over E_4+2\chi E_6+\chi^2{E_4}^2}\right)^{1/4}.}
The natural coordinate, $\phi$, on the moduli space of the
gauge theory \SW\ is given by
\eqn\phieq{
\phi={\pi\over\b}\int {du\over u^{1/2}}
\left({\tE_4\over E_4+2\chi E_6+\chi^2{E_4}^2}\right)^{1/4}.}
In the limit that $m$ approaches zero, one has $\ttau\to\tau$ and
$\phi\to u^{1/2}$, thus from \phieq, $\b=2\pi$.  We can expand the
integral in powers of $u$, where in the large $u$ limit, we would
find all half integer powers, except for an integration constant.
This integration constant
is set to zero so that the $Z_2$ invariance under $\phi\to
-\phi$, $u\to u$ is preserved.

We now explore the modular properties of these various terms. Under
the transformation $\tau\to{a\tau+b\over c\tau+d}$, the $E_k$
transform as $E_k\to (c\tau+d)^k E_k$. Hence the curve in \curverw\
transforms nicely under these modular transformations if we also
assume that $u\to (c\tau+d)^2 u$, $m\to m$. Hence $\chi$ has the
transformation $\chi\to (c\tau+d)^{-2}\chi$.

By comparing \ellcurve\ and \curverw\ we can find another relation,
\eqn\Erel{
{{\tE_4}^3\over{\tE_6}^2}=
{\left(E_4+2\chi E_6+\chi^2{E_4}^2\right)^3\over
(E_6+3{E_4}^2\chi+
3E_4E_6\chi^2+(2{E_6}^2-{E_4}^3)\chi^3)^2}
}
{}From \Erel, and the fact that $\ttau\to\tau$ as $ m \to 0$,
we immediately learn that the modular transformation on
$\tau$ induces the transformation $\ttau\to {a\ttau+b\over
c\ttau+d}$.  It then follows that the difference $\ttau-\tau$
transforms as:
\eqn\taudtr{
\ttau-\tau\to (c\ttau+d)^{-1}(c\tau+d)^{-1}(\ttau-\tau) .}

Next consider the dual coordinate $\phi_D$ which is the
integral over $u$ of the dual period $\om_D = \ttau \om$:
\eqn\phiDeq{
\phi_D={1\over2}\int {du\over u^{1/2}}
\left({\tE_4\over E_4+2\chi E_6+\chi^2{E_4}^2}\right)^{1/4}
\ttau.}
If we now consider $\phi_D-\tau\phi$, then from \phiDeq, \phieq,
\taudtr\ and the action of the modular transformation on $u$, we have
\eqn\phidtr{
\phi_D-\tau\phi\to (\phi_D-\tau\phi)(c\tau+d)^{-1}.}
Hence we may infer that the $u$ expansion of $\phi_D-\tau\phi$ is
\eqn\pduexp{
\phi_D-\tau\phi=\sum_{n=0}p_{2n}(\tau){m^{2n+2}\over u^{1/2+n}}
}
where the terms $p_{2n}(\tau)$ are modular forms of weight $2n$.
Hence, they are composed of rational functions of $E_4$ and $E_6$.
Indeed, for $\chi$ small, one can see explicitly from \phiDeq\ that
the $p_{2n}(\tau)$ are polynomials in $E_4$ and $E_6$

However, in order to find the instanton expansion for the free energy
and
the coupling,  we actually want the expansion of $\phi_D$ in powers
of
$\phi$, not
$u$.  Moreover, $u$ is a somewhat arbitrary parameter for an elliptic
curve, while $2\phi$ is the physical mass of the gauge bosons. But,
it
is clear from \phieq\ that an expansion of $\phi_D$ in terms of
$\phi$
will not have coefficients that are modular forms;  under the modular
transformation, $\phi$ transforms to $c\phi_D+d$.

Luckily, this loss of modular invariance is actually quite mild and
in
fact can be calculated precisely.  Furthermore, the breaking of the
modular invariance is described by a modular anomaly equation which
can be used to generate a recursion relation.

Let $H=\ttau-\tau$ and define a new expression $h$,
\eqn\heq{
h={H\over 1-{2\pi i\over12}E_2 H},
}
where $E_2$ is the regulated Eisenstein series of weight two and
which
is also given by  $E_2={24\over 2\pi
i}\partial_\tau\log(\eta(\tau))$.
The function $E_2$ is
not quite a modular form, under a modular transformation it generates
an extra piece, with
\eqn\EIIeq{
E_2\left({a\tau+b\over c\tau+d}\right)=(c\tau+d)^2\left(E_2(\tau)+
{12\over2\pi i}{c\over c\tau+d}\right).
}
Using \EIIeq\ and \phidtr\ we find that under a modular
transformation,
$h\to (c\tau+d)^{-2}h$.   Therefore, the $u$ expansion of $h$ is
comprised
solely of modular forms and so has no $E_2$ terms in the
expansion.
Thus,
\eqn\Heq{
{\del H\over\del E_2}={\del\over\del E_2}\left({h\over1+{2\pi
i\over12}E_2 h}
\right)=-{2\pi i\over12}H^2
}

Next consider the term
\eqn\Integrand{
(\tE_4/\bE_4)^{1/4}(\ttau-\tau),}
 where $\bE_4$
is given by
\eqn\barEeq{
\bE_4=E_4+2\chi E_6+\chi^2 E_4^2.
}
The expression in \Integrand\ is a modular function of weight $-2$,
therefore its $u$ expansion
is independent of $E_2$.  Using this fact and \Heq\ we find that
\eqn\Geq{
{\del \over\del E_2}\left({\tE_4\over\bE_4}\right)^{1/4}={2\pi
i\over12}
\left({\tE_4\over\bE_4}\right)^{1/4}(\ttau-\tau).
}

We now want to treat $\tau$ and $\phi$ as the independent variables.
 As
a function of $\tau$ and $\phi$, $u$ can have some $E_2$ dependence.
To
this end let us act on \phieq\ with $\del_{E_2}$, giving the equation
\eqn\phieqdEII{
0={1\over2}{\del u\over \del E_2} {1\over u^{1/2}}
\left({\tE_4\over\bE_4}\right)^{1/4}+\half\int {du\over u^{1/2}}
{\del \over\del E_2}\left({\tE_4\over\bE_4}\right)^{1/4}.}
The derivative with respect to $E_2$ inside the integral on the
right-hand side
refers
to the {\it explicit} $E_2$ dependence in the $u$ expansion of
$(\tE_4/\bE_4)^{1/4}$.  Hence, using \Geq\ and \phiDeq\
in \phieqdEII\ one finds
\eqn\dudE{
{\del u\over \del E_2} =-{2\pi
i\over12}\left({\tE_4\over\bE_4}\right)^{-1/4}
(\phi_D-\tau\phi)
}
 We can also act on \phieq\ with $\del_\phi$, which
gives the equation
\eqn\phieqdphi{
1={1\over2}{\del u\over \del \phi} {1\over u^{1/2}}
\left({\tE_4\over\bE_4}\right)^{1/4}.
}
Therefore, from \dudE\ we have
\eqn\dudEII{
{\del u\over \del E_2} =-{2\pi i\over12}{\del u\over \del \phi}
(\phi_D-\tau\phi)
}
Using \dudEII\ in the two relations
\eqn\dEdphi{\eqalign{
\del_{E_2}(\phi_D-\tau\phi)& = {\del u\over \del
E_2}\del_u(\phi_D-\tau\phi)\cr
\del_{\phi}(\phi_D-\tau\phi)& = {\del u\over \del
\phi}\del_u(\phi_D-\tau\phi)
}}
we derive the equation
\eqn\phiDd{
\del_{E_2}(\phi_D-\tau\phi)=-{2\pi
i\over24}\del_{\phi}(\phi_D-\tau\phi)^2.
}
Since $\phi_D=\del_\phi\cF$ where $\cF$ is the pre-potential, \phiDd\
can
be integrated, giving
\eqn\cFd{
\del_{E_2}\ctF=-{2\pi
i\over24}\left(\del_{\phi}\ctF\right)^2+C(\tau),
}
where $\ctF=\cF-\ha\tau\phi^2$ and $C(\tau)$ is a $\phi$ independent
term.

The equation in \cFd\ has exactly the same form as an equation in
\MNWII,
where recursion relations were derived for the instanton expansion of
the $E_8$ non-critical string.  However, the expressions for $\ctF$ are
completely different since the expansion parameters are
different and the perturbative contribution for the noncritical string is
trivial.  

The key point is that $\ctF=0$ if $m=0$.  Therefore, the recursion
relation we derive should relate the higher power terms in the $m$
expansion
of $\ctF$ to the lower power terms.  Since only even powers of $m$
appear,
the series expansion for $\ctF$ has the form
\eqn\seriesF{
\ctF={1\over8\pi i}f_1(\tau)\ m^2\log(2\phi/m^2)^2
-{1\over4\pi i}\sum_{n=2}~{f_n(\tau)\over (2n-2)}
{m^{2n}\over(2\phi)^{2n-2}},
}
up to a $\phi$ independent piece that does not effect the physics.
We have included a factor of $2$ in front of $\phi$ for later
convenience.  The coefficients $f_n(\tau)$ are polynomials of $E_2$,
$E_4$
and $E_6$, with  weight $2n-2$, so that the total weight of $\ctF$
is zero.  If we plug the expansion of $\ctF$ in \seriesF\ into \cFd,
we
find the recursion relation
\eqn\recrel{
\del_{E_2}f_n=\left({n-1\over6}\right)\sum_{m=1}^{n-1}f_mf_{n-m}.
}
Hence, up to the $E_2$ independent terms, if we know $f_1$, then we
can generate the whole instanton expansion.

\subsec{The instanton expansion from the recursion relation}

The first term $f_1$ in \seriesF\
can be found from the one-loop piece of $\ctF$.
The perturbative piece also contributes to the higher $f_n$, but no
instanton term contributes to $f_1$.  The perturbative pre-potential
is
given by
\eqn\prepert{
\ctF_{pert}={-1\over 8\pi i}\left((2\phi)^2\log(2\phi)^2-
\ha\left((2\phi+m)^2\log(2\phi+m)^2+
(2\phi-m)^2\log(2\phi-m)^2\right)\right).
}
The first few terms in its $m$ expansion are
\eqn\pertm{
\ctF_{pert}={1\over8\pi
i}\left(m^2\log(2\phi)^2-{1\over6}{m^4\over(2\phi)^2}-
{1\over30}{m^6\over(2\phi)^4}\right)+{\rm O}(m^8)
}
Therefore, $f_1=1$, and so from \recrel,  $f_2={1\over6}E_2$.
 The expansions of the $E_n$ are given by
\eqn\Eexp{\eqalign{
E_2&=1-24\sum_{n=1}\s_1(n)q^n\cr
E_4&=1+240\sum_{n=1}\s_2(n)q^n\cr
E_6&=1-504\sum_{n=1}\s_3(n)q^n,
}}
where $q=e^{2\pi i\tau}$ and $\s_k(n)={\sum\atop d|n}d^k$.
Comparing $f_2$ and \pertm, we see that the coefficient in front of
$E_2$ is
correctly normalized.  The next term in the series is $f_3$, which
satisfies
$f_3={1\over18}{E_2}^2+\a E_4$
We can determine $\a$ by comparing to the $m^6$ term in \prepert.
Matching this to the  $q$ independent piece in $f_3$, we find that
$\a=1/90$.
The next three terms can be found in a similar manner, and are given
by
\eqn\nextIII{\eqalign{
f_4&={5\over216}{E_2}^3+{1\over90}E_2E_4+{11\over 7560}E_6\cr
f_5&={7\over648}{E_2}^4+{7\over810}{E_2}^2E_4+{19\over 22680}{E_4}^2+
{11\over5670}E_2E_6\cr
f_6&={7\over
1296}{E_2}^5+{1\over162}{E_2}^3E_4+{17\over11340}E_2{E_4}^2
+{11\over6048}{E_2}^2E_6+{37\over142560}E_4E_6.
}}

The perturbative expansion does not give enough information to
compute
$f_7$ and beyond.  This is because at $f_7$, there are two
independent
modular forms of weight 12, generated by ${E_4}^3$ and ${E_6}^2$.  So
we need another piece of information to set their coefficients.  In
order to
proceed, let us
look at the $q$ expansions for the first six terms.  The expansions
are
\eqn\fqexp{\eqalign{
f_1& =1\cr
f_2&={1\over6}-4q-12q^2-16q^3+28q^4+{\rm O}(q^5)\cr
f_3&={1\over15}+48q^2+256q^3+720q^4+{\rm O}(q^5)\cr
f_4&={1\over28}-30q^2-960q^3-6570q^4+{\rm O}(q^5)\cr
f_5&={1\over45}+{3584\over3}q^3+24864q^4+{\rm O}(q^5)\cr
f_6&={1\over66}-480q^3-43020q^4+{\rm O}(q^5)\cr
}}
An obvious feature is that after the constant term, the $q$ expansion
starts at $q^2$ for $f_3$ and $f_4$ and at $q^3$ for $f_5$ and $f_6$.

We can understand the presence of these ``gaps''
 by considering the $N=2$ pure gauge limit when
the mass of the adjoint scalar is taken to infinity.  At the same
time
we also must take $\tau$ to imaginary infinity in order that the
resulting
cut off is finite.  In the limit of large $m$ and $q$ the
perturbative
part of the pre-potential is to leading order
\eqn\prepertlim{
\cF\approx\ha\tau\phi^2+{1\over2\pi
i}\phi^2\log\left(m^2/\phi^2\right)=
{1\over2\pi i}\phi^2\log\left(m^2q^{1/2}/\phi^2\right)
}
Hence, in order to keep things finite, we should scale the cutoff as
$\Lambda^2=m^2q^{1/2}$.  If we now look at the contributions to the
instantons, we see that any term of the form $(m^4q)^s m$ will
diverge
in this limit.  Thus, if we want to avoid such terms, we require that
the
$q$ expansion of $f_{2n}$ starts at $q^n$ and that of $f_{2n+1}$
starts
at $q^{n+1}$.

We can use the gaps to fix the coefficients of the modular
forms.
In fact, the presence of the gaps has made the system overdetermined
and provides a nontrivial check on our formalism.  The gaps have also
obviated the need for the $m$ expansion of $\ctF_{pert}$ beyond the
leading order
term.  So, to find $f_7$ we can adjust the coefficients of ${E_4}^3$
and
${E_6}^2$ such that there is no $q$ and $q^2$ term in the expansion.
We
can then check that everything is consistent by making sure that
there
is no $q^3$ term either.

Using {\it Mathematica} we were able to generate the first 48 terms.
The
next six terms in the expansion are
$$\eqalign{
f_7&={11\over3888}{E_2}^6+{11\over2592}{E_2}^4E_4+
{1199\over680400}{E_2}^2{E_4}^2+{2281\over23351328}
{E_4}^3+{121\over81648}
{E_2}^3E_6\cr
&\qquad
+{4127\over7484400}E_2E_4E_6+{3313\over145945800}{E_6}^2\cr
f_8&={143\over93312}{E_2}^7+{1001\over349920}{E_2}^5E_4+
{377\over218700}
{E_2}^3{E_4}^2+{62459\over250192800}E_2{E_4}^3+
{1573\over1399680}{E_2}^4E_6\cr
&\qquad+
{56797\over76982400}{E_2}^2E_4E_6+{7907\over159213600}{E_4}^2E_6+
{43151\over 778377600}E_2{E_6}^2
}$$
\eqn\nextsix{\eqalign{
f_9&={715\over839808 }{E_2}^8+{1001 \over 524880}{E_2}^6E_4+
{533 \over 349920}{E_2}^4{E_4}^2+
{146057 \over 375289200}{E_2}^2{E_4}^3+{107803 \over
7695160704}{E_4}^4\cr
&\qquad+
{143 \over 174960}{E_2}^5E_6+{1537 \over 1924560}{E_2}^3E_4E_6
+{54928 \over383107725 }{E_2}{E_4}^2E_6+
{202831 \over 2451889440}{E_2}^2{E_6}^2\cr
&\qquad
+{2947 \over327175200 }E_4{E_6}^2\cr
f_{10}&={2431 \over5038848 }{E_2}^9 +{221 \over 174960}{E_2}^7{E_4}
+{221 \over174960 }{E_2}^5{E_4}^2
+{6300863 \over13135122000 }{E_2}^3{E_4}^3 \cr
&\qquad
+{9656989 \over208410602400 }{E_2}{E_4}^4
+{2431 \over4199040 } {E_2}^6{E_6}
+{82501 \over107775360 } {E_2}^4{E_4}{E_6} \cr
&\qquad
+{1890281 \over7662154500 }{E_2}^2{E_4}^2{E_6}
+{12343193 \over 1234223827200}{E_4}^3{E_6}
+{1438081 \over14711336640 }{E_2}^3{E_6}^2 \cr
&\qquad
+{34431581 \over1190917728000 } {E_2}{E_4}{E_6}^2
+{118163 \over 203443488000} {E_6}^3\cr
f_{11}&={4199 \over 15116544 }{E_2}^{10} + {4199 \over5038848 }
{E_2}^8{E_4}+
{44251 \over44089920 }{E_2}^6{E_4}^2
+ {129785599 \over 252194342400}{E_2}^4{E_4}^3 \cr
&\qquad+
{1345155521 \over 15005563372800}{E_2}^2{E_4}^4
+ {233831 \over 103205684736}{E_4}^5 +
{3553 \over 8817984}{E_2}^7{E_6} \cr
&\qquad
+ {329137 \over484989120 }{E_2}^5{E_4}{E_6} +
{72801977 \over220670049600 }{E_2}^3{E_4}^2{E_6}
+ {5185188971 \over142552852041600 }{E_2}{E_4}^3{E_6}\cr
&\qquad  +
{ 11850547\over117690693120 }{E_2}^4{E_6}^2
 + { 407489447\over 7502781686400}{E_2}^2{E_4}{E_6}^2  +
{94033729 \over33541847539200 }  {E_4}^2{E_6}^2\cr
&\qquad
 + {79550227 \over38878050556800 } {E_2}{E_6}^3 \cr
f_{12}&={29393 \over 181398528}{E_2}^{11}
+{24871 \over 45349632}{E_2}^{9}{E_4} +
{3553 \over4592700 }{E_2}^7{E_4}^2
 +{4951267 \over9825753600 }{E_2}^5{E_4}^3\cr
&\qquad
 + {24269491 \over 182697606000}{E_2}^3{E_4}^4
+{40781021 \over4443205777920 } {E_2}{E_4}^5
+ {39083 \over 141087744}{E_2}^8{E_6}\cr
&\qquad
+{191539 \over335923200 } {E_2}^6{E_4}{E_6}
+ {26069273 \over68780275200 }{E_2}^4{E_4}^2{E_6}
+{16939036637 \over 222160288896000}{E_2}^2{E_4}^3{E_6} \cr
&\qquad
+ { 1154742773\over552204732119040 }{E_4}^4{E_6}
 +{8693849 \over91707033600 }{E_2}^5{E_6}^2 \cr
&\qquad+
{91430249 \over1169264678400 }{E_2}^3{E_4}{E_6}^2
 +{ 379916683957\over 34212684489984000}{E_2}{E_4}^2{E_6}^2 \cr
&\qquad
+ {7758245077 \over 1866146426726400}{E_2}^2{E_6}^3
 +{71140783 \over195048767232000 }{E_4}{E_6}^3
}}

We have already seen that the term $f_{2n}$ starts its series
expansion
at $q^n$ and $f_{2n+1}$ starts its series expansion at $q^{n+1}$.
Therefore,
in order to compute the entire $n$ instanton contribution to $\cF$,
we
need to compute the series expansions up to the $q^n$ term of $f_i$
for
$i$ from $1$ to $2n$.  Hence, with 48 terms we can compute the first
24
terms in the instanton expansion.  The first eight are
\eqn\instVIII{\eqalign{
\cF_1&={1\over2\pi i}{m^4\over(2\phi)^2}q\cr
\cF_2&={1\over2\pi i}\left(3{m^4\over(2\phi)^2}
-6{m^6\over(2\phi)^4}+{5\over2}{m^8\over(2\phi)^6}\right)q^2\cr
\cF_3&={1\over2\pi
i}\left(4{m^4\over(2\phi)^2}-32{m^6\over(2\phi)^4}+
80{m^8\over(2\phi)^6}-{224\over3}{m^{10}\over(2\phi)^8}+
24{m^{12}\over(2\phi)^{10}}\right)q^3\cr
\cF_4&={1\over2\pi
i}\biggl(7{m^4\over(2\phi)^2}-90{m^6\over(2\phi)^4}+
{1095\over2}{m^8\over(2\phi)^6}-1554{m^{10}\over(2\phi)^8}+
2151{m^{12}\over(2\phi)^{10}}\cr
&\qquad
-1430{m^{14}\over(2\phi)^{12}}+
{1469\over4}{m^{16}\over(2\phi)^{14}}\biggr)q^4\cr
\cF_5&={1\over2\pi
i}\biggl(6{m^4\over(2\phi)^2}-192{m^6\over(2\phi)^4}+
{2144}{m^8\over(2\phi)^6}-12096{m^{10}\over(2\phi)^8}+
{187056\over5}{m^{12}\over(2\phi)^{10}}\cr
&\qquad
-65472{m^{14}\over(2\phi)^{12}}+
{323232\over5}{m^{16}\over(2\phi)^{14}}-33600{m^{18}
\over(2\phi)^{16}} +
{35768\over5}{m^{20}\over(2\phi)^{18}}\biggr)q^5\cr
\cF_6&={1\over2\pi
i}\biggl(12{m^4\over(2\phi)^2}-360{m^6\over(2\phi)^4}+
{6210}{m^8\over(2\phi)^6}-58016{m^{10}\over(2\phi)^8}+
{314016}{m^{12}\over(2\phi)^{10}}\cr
&\qquad
-1033120{m^{14}\over(2\phi)^{12}}+
{2114840}{m^{16}\over(2\phi)^{14}}-2698080{m^{18}\over(2\phi)^{16}}+
{6249064\over3}{m^{20}\over(2\phi)^{18}}\cr
&\qquad
-890112{m^{22}\over(2\phi)^{20}}+
{161588}{m^{24}\over(2\phi)^{22}}\biggr)q^6\cr
\cF_7&={1\over2\pi
i}\biggl(8{m^4\over(2\phi)^2}-576{m^6\over(2\phi)^4}+
{14880}{m^8\over(2\phi)^6}-207168{m^{10}\over(2\phi)^8}+
{1727856}{m^{12}\over(2\phi)^{10}}\cr
&\qquad
-9109760{m^{14}\over(2\phi)^{12}}+
{219699584\over7}{m^{16}\over(2\phi)^{14}}-71919360
{m^{18}\over(2\phi)^{16}}+
{109991904}{m^{20}\over(2\phi)^{18}}\cr
&\qquad-
{774893568\over7}{m^{22}\over(2\phi)^{20}}+
{70299264}{m^{24}\over(2\phi)^{22}}-
{25518592}{m^{26}\over(2\phi)^{24}}+
{28244800\over7}{m^{28}\over(2\phi)^{26}}\biggr)q^7\cr
\cF_8&={1\over2\pi
i}\biggl(15{m^4\over(2\phi)^2}-930{m^6\over(2\phi)^4}+
{62635\over2}{m^8\over(2\phi)^6}-605934{m^{10}\over(2\phi)^8}+
{7192017}{m^{12}\over(2\phi)^{10}}\cr
&\qquad
-55338690{m^{14}\over(2\phi)^{12}}+
{1146161679\over4}{m^{16}\over(2\phi)^{14}}-
1022959770{m^{18}\over(2\phi)^{16}}+
{2552909057}{m^{20}\over(2\phi)^{18}}\cr
&\qquad -
{4471880166}{m^{22}\over(2\phi)^{20}}+
{10919211219\over2}{m^{24}\over(2\phi)^{22}}-
{4541726970}{m^{26}\over(2\phi)^{24}}\cr
&\qquad+
{2451618975}{m^{28}\over(2\phi)^{26}}-
{773708598}{m^{30}\over(2\phi)^{28}}+
{866589165\over8}{m^{32}\over(2\phi)^{30}}\biggr)q^8
}}
The two instanton expression matches the result of D'Hoker and Phong
for
$SU(2)$ \DP.

We can directly find the instanton expansion for $N=2$ $SU(2)$ Super
Yang-Mills by taking the limit $m\to\infty$, $m^4q=\Lambda^4$. The
surviving terms in the instanton expansion are those with the highest
power of $m$. When can readily check that these terms match those
found by Matone \Matone.

A striking feature of the $\cF_n$ terms in \instVIII\ is that after
taking a $\tau$ derivative, the coefficients of the $m^2$ expansions
are integers. This suggests that the instanton expansion is computing
toplological invariants, and the natural guess is that they are
Euler numbers for instanton moduli spaces.

\subsec{The coupling near the $U(1)$ singularity}

The instanton expansion also has an interesting behavior as $m$
approaches
$2\phi$.  At this point a charged BPS state is massless,  hence
the perturbative piece has a log singularity,
Let us consider the expansion of the coupling as a series in
$(1-m^2/(2\phi)^2)$.  The full perturbative contribution of this
effective $U(1)$ theory is
\eqn\taueff{
\tau_{eff}={1\over 2\pi i}\log\left(1-{m^2\over(2\phi)^2}
\right)^2 +{1\over 2\pi i}\log(C(\tau)), }
where $C(\tau)$ is a function to be determined.  The nonperturbative
pieces of
the effective $U(1)$ theory appear as higher powers of
$(1-m^2/(2\phi)^2)$.
To find $C(\tau)$, consider the expression
${{\tE_4}^3-{\tE_6}^2\over{\tE_4}^3}$.  Using \Erel, \Eisen\ and
\barEeq,
we find that
\eqn\Etexp{
{{\tE_4}^3-{\tE_6}^2\over{\tE_4}^3}={{E_4}^3-{E_6}^2\over{\bE_4}^3}
\prod_{i=1}^3(1-3\chi e_i)^2.
}
The hypermultiplet is massless if $\chi=1/(3e_1)$, which corresponds
to $u=m^2e_1/4$.  At this singularity
\eqn\singeq{
\tq=0,\qquad\qquad e_1^2\bE_4=(e_1-e_2)^2(e_1-e_3)^2,
}
therefore, near the singularity we can approximate \Etexp\ as
\eqn\Etexpapp{
1728\tq={1728\eta^{24}(\tau){e_1}^2\over (e_1-e_2)^4(e_1-e_3)^4}
\left(1-{m^2e_1\over4u}\right)^2+{\rm O}(\tq^2).
}

However, we want to express \Etexpapp\ in terms of $\phi$ and not
$u$.  Since
the singularity occurs at $\phi^2=m^2/4$, we can approximate $\phi^2$
as
\eqn\phiapp{
\left(1-{m^2\over4\phi^2}\right)=\a\left(1-{m^2e_1\over4u}\right)+
{\rm O}\left(\left(1-{m^2e_1\over4u}\right)^2\right).
}
In order to determine
$\a$ in \phiapp, consider the expression for $\phi$ in \phieq, with
$\b=2\pi$.
Taking a derivative with respect to $u$, we have
\eqn\phider{
{\partial\phi\over\partial u}\biggr\vert_{u=m^2e_1/4}={1\over
2u^{1/2}}
\left({\tE_4\over\bE_4}\right)^{1/4}\Biggr\vert_{u=m^2e_1/4}=
{1\over m\sqrt{(e_1-e_2)(e_1-e_3)}}
}
Taking a $u$ derivative on \phiapp, we have
\eqn\phiappd{
{m^2\over2\phi^3}{\partial\phi\over\partial
u}\biggr\vert_{u=m^2e_1/4}=
\a {m^2 e_1\over 4u^2}\biggr\vert_{u=m^2e_1/4}={4\a\over m^2e_1}
}
Thus, we have
\eqn\aleq{
\a={e_1\over\sqrt{(e_1-e_2)(e_1-e_3)}}.
}
and therefore, $C(\tau)$ is
\eqn\Cteq{
C(\tau)={\eta^{24}(\tau)\over (e_1-e_2)^3(e_1-e_3)^3}=
{\eta^{24}(2\tau)\over\eta^{24}(\tau)},
}
where we have used the fact that $e_1-e_2={\vartheta_3}^4$ and
$e_1-e_3={\vartheta_4}^4$.  We can now explicitly check that the
instanton
expansions in \instVIII\ are consistent with \Cteq.  

As was mentioned before,  the log of
the right-hand side of \Cteq\ looks like the holomorphic part of a string 
threshhold correction, which can also be expressed in terms of $F_1$,
the genus one partition function of
a topological field theory coupled to topological gravity\BCOV.
If the target space is a two torus with Kahler modulus $t$, 
then $F_1$ satisfies
\eqn\topamp{
\del_t F_1={2\over2\pi i}\del_t\log \eta\left(\exp(2\pi it)\right)
}
The right-hand side is the generator for the number of inequivalent maps
of a world-sheet torus to a target space torus.  Hence, the $\tau$
derivative of  ${1\over12}\tau_{eff}$ in \taueff\
seems to count the number of maps from a torus into a torus with modulus 
$2\tau$ minus the number of maps into a torus with modulus $\tau$.
This is made even more suggestive if one considers the behavior of the
Donagi-Witten curve at the singular point $m^2=4\phi^2$, where a genus
two surface degenerates into a torus with Teichmuller parameter $2\tau$.

\newsec{Recursion relations for $SU(n)$}

It is relatively straightforward to generalize our results
to other gauge groups, and to $SU(n)$ in particular.  To do
this we first recall some of the essential features of
the quantum effective action of softly broken $N=4$
supersymmetric $SU(n)$ gauge theories \RDEW.

\subsec{The $SU(n)$ quantum effective action}

The relevant Riemann surface is defined as an $n$-sheeted
foliation over a torus.  That is, one introduces the standard
Weierstrass torus, with modulus $\tau$:
\eqn\weierTor{ y^2 ~=~ 4~(x - e_1(\tau))~(x - e_2(\tau))~
(x - e_3(\tau)) \ ;}
and defines the Riemann surface over it via:
\eqn\foliate{F_n(t,x,y; u_j) ~\equiv~ \sum_{\ell = 0}^n ~
u_{n - \ell} ~ P_{\ell} (t,x,y) ~=~ 0 \ ,}
where $u_0 = 1$.  The $P_\ell$ are quasi-homogeneous polynomials
of weight $\ell$, (where $t,x,y$ are assigned weights $1,2,3$), and
are uniquely defined by requiring $P_\ell \sim t^\ell + \dots$ as
$t \to \infty$, and by specifying their factorization properties in
the limit  $x,y,t \to \infty$ \RDEW.  We have deviated slightly from
\RDEW\ in that we have taken $u_1 \not = 0$.  We have done this
for later mathematical convenience, and it may be thought of as
as adding an extra $U(1)$ factor, converting the action to that of
a  $U(n)$ gauge theory.  It should be noted that the surface
\foliate\ has genus $n$, while $SU(n)$ has rank $n-1$.  The
reason for promoting the gauge group to $U(n)$ is that we
want the genus of the surface to equal the rank of the gauge
group.

The Seiberg-Witten differential for these models is
given by \refs{\RDEW\HIAM{--}\EMa}:
\eqn\lswSUn { \lambda_{SW} ~=~ t ~{dx \over y} ~=~ t~d\xi \ ,}
where $\xi$ is the standard ``flat'' coordinate on the base torus.
The periods of this differential are thus
\eqn\phisSUn{\phi_i ~=~ \oint_{a_i}~\lambda_{SW} \ , \qquad
\phi_{D,i} ~=~ \oint_{b_i}~\lambda_{SW} \ ,  \qquad i = 1,\dots,n.}
There are thus $n$ of each of the $\phi_i$ and the $\phi_{D,i}$.
However, let $t_j, j = 1,\dots,n$ denote the roots of \foliate.
Since $P_n$ contains no term
$t^{n-1}$ term, it follows that $u_1 = \sum t_j$, and hence
\eqn\sumphis{\sum_{i=1}^n ~\phi_i ~=~ u_1~ \oint_{a}~ {dx \over y}
\ , \qquad \sum_{i=1}^n ~\phi_{D,i} ~=~ u_1~ \oint_{b}~{dx \over y}
\ ,  \qquad i = 1,\dots,n \ ,}
where $a$ and $b$ are the cycles of the base torus.
Thus the sum of the $\phi_i$ and $\phi_{D,i}$ are simply the
standard periods of the base torus multiplied by $u_1$.  This
sum gives the effective action of the $U(1)$ factor in $U(n)$,
and the fact that it depends trivially on the $u_j$ and the
base torus reflects the triviality of the effective action of
a pure $U(1)$ theory (with no coupling to charged matter).
Setting $u_1 =0$ as in  \RDEW\  means that the $\phi_j$ and
$\phi_{D,j}$ are not all independent (and indeed this
constraint projects one onto the rank $n-1$ Prym variety of
interest).  We have really introduced $u_1$ to avoid having
to deal with this constraint and thereby avoid littering the
discussion below with projection operators.

The dependence on the hypermultiplet mass, $m$, is implicit  in the
discusson above:  The $u_j$ are the invariants of the Higgs vevs
divided by $m^j$.  To restore the mass dependence explicitly,
one rescales $t \to t/m$, replaces $u_j \to u_j/m^j$, and multiplies
$F_n$ by $m^n$.  The net effect is the same as replacing
$x \to m^2 x, y \to m^3 y$ in $F_n$.  The $m = 0$ limit is now
clear:  One has $P_\ell = t^\ell$, and the foliation over
over the base is trivial.  The Riemann surface is $n$ disconnected
copies of the base torus \weierTor.  The Seiberg-Witten differential
on the $j^{\rm th}$ copy is $t_j d\xi$, where $t_j$ is the
$j^{\rm th}$ root of \foliate.  Note that $t_j$ is constant over the
base, and so the periods of $\lambda_{SW}$ are simple multiples
of the periods of the base torus.

\subsec{The $SU(n)$ recursion relation}

As in section 2, we imagine expanding the periods of the
Seiberg-Witten differential about $m=0$, in a series in $m$
and inverse powers of $\phi_i$, or more precisely
$\phi_i - \phi_j$.  Our goal is to generalize
\phiDd\ and \cFd, and the key to doing it
is to establish the analog of \Heq\ and establish and
use the modular properties of appropriate generalization of
$\phi_D - \tau \phi$.

Let $\zeta_j$ be a canonically normalized
basis of holomorphic differentials on the Riemann surface
\foliate\ at a general value of $m$, {\it i.e.}
\eqn\permat{\oint_{a_i}~\zeta_j ~=~ \delta_{ij} \ , \qquad
\oint_{b_i}~\zeta_j ~=~ \widetilde \Omega_{ij} \ ,  \qquad
i,j = 1,\dots,n .}
The period matrix is thus $({\cal I}, \widetilde \Omega)$
where ${\cal I}$ is the $n \times n$ identity matrix.
Let $({\cal I}, \Omega)$ be the the corresponding period matrix
at $m=0$. Since the surface at $m=0$ is $n$ disconnected copies of
the
the base, one has:
\eqn\Omform{\Omega ~=~ \tau~{\cal I} \ .}
An element ${\cal M}$ of the group $Sp(2g,\ZZ)$ acts upon the
cycles, and hence on the periods, by left multiplication:
\eqn\Spaction{\left(\matrix{b_j \cr a_i} \right)  ~\to~
{\cal M}~\left(\matrix{b_j \cr a_i}\right)  \ , \qquad
\left(\matrix{\phi_{D,j} \cr \phi_i} \right) ~\to~ {\cal M}~
\left(\matrix{\phi_{D,j}\cr \phi_i} \right)  \ .}
If one writes ${\cal M}$ in terms of its $n \times n$ blocks
then its action on $\widetilde \Omega$ may be written:
\eqn\projaction{{\cal M} ~=~ \left(\matrix{A & B \cr C & D} \right) \
,
\qquad  \widetilde \Omega ~\to~ (~A~\widetilde \Omega ~+~ B~)~
(~C~\widetilde \Omega ~+~ D~)^{-1} \ .}
The submatrices of ${\cal M}$ must satisfy:
\eqn\sympcond{A^T C = C^T A \ ; \quad B^T D = D^T B   \ ; \quad
A^T D ~-~ C^T B  ~=~ D^T A ~-~ B^T C  ~=~ {\cal I} \ .}

As in section 2, introduce a matrix $H = \widetilde \Omega - \Omega$.
Under the modular transformation, ${\cal M}$, one has:
\eqn\Htransf{\eqalign{ H ~\to~ & (~\Omega~C^T ~+~ D^T~)^{-1} ~H~
(~C~\widetilde  \Omega ~+~ D~)^{-1} \ , \cr
{}~=~  & (~\Omega~C^T ~+~ D^T~)^{-1} ~H~ (~C~H ~+~
C~\Omega ~+~ D~)^{-1} \ .}}
Define a matrix, $h$, via
\eqn\hmatdefn{h ~=~ H~\big(1 ~-~ {2 \pi i \over 12}~E_2(\tau)~ H
\big)^{-1} \ .}
This does not behave well under a general $Sp(2g,\ZZ)$
transformation,
but  we now specialize to the $SL(2,\ZZ)$ subgroup of $Sp(2g,\ZZ)$
induced via modular transformations of the base torus, {\it
i.e.} we take $(A,B,C,D) = (a {\cal I}, b {\cal I}, c {\cal I},
d {\cal I})$.  One then finds that all the entries of the matrix $h$
are ($SL(2,\ZZ)$) modular forms of weight $-2$, {\it i.e.} one
has $h \to (c \tau + d)^{-2} h$.  As before one can write the entries
of $h$ in terms of the modular forms $E_4$ and $E_6$, and hence
$H$ may be written in terms of $E_2, E_4$ and $E_6$, and once again
\Heq\ is satisfied, where $H$ is now a matrix.

{}From \Spaction\ and \sympcond\ it is easily seen that the column
vectors $\phi_D - \Omega \phi$ transform under $Sp(2g,\ZZ)$ as
\eqn\phistrf{ \phi_D ~-~ \Omega~\phi ~\to~ (~\Omega~C^T ~+~
D^T ~)^{-1}~ \big(~\phi_D ~-~ \Omega~\phi~\big) \ .}
It follows that under the $SL(2,\ZZ)$ subgroup, each element
of $\phi_D - \Omega \phi$ transforms as a modular form of
weight $-1$, exactly as in \phidtr.

To analyze expansions about $m=0$, we need to consistently
assign formal modular weights under $SL(2,\ZZ)$.  First, because
$x$ is a Weierstrass function on the torus, it follows that
$x$ has weight $2$, and $y$  has weight $3$.  This means that
$t$ must be assigned a formal weight of $1$, and the $u_j$ are
to be given a formal weight $j$ for the defining equation of
the Riemann surface \foliate\ to be modular invariant.
Let $v_j$ denote the roots of $F_n = 0$ at $m=0$, {\it i.e.}
the $v_j$ are the roots of $t^n + \sum_{j=1}^n  u_j t^{n-j} = 0$.
The $u_j$ are thus invariant polynomials in the $v_j$.  Moreover,
since $SL(2,\ZZ)$ acts on each leaf of the foliation separately,
it does not permute the $v_j$, and so the $v_j$ thus have a
formal modular weight of $1$.
Suppose that one expands $\phi_{D,i} - \tau \phi_i$ into a
Laurent series of the form:
\eqn\Lseries{\phi_{D,i} - \tau \phi_i ~=~  \sum_{k = 0}^\infty ~
\sum_{\alpha} ~G_{\alpha,k} (\tau) ~{m^{k+1}  \over  Q_{\alpha,k}
(v_\ell)} \ ,}
where the $Q_{\alpha,k}(v_\ell)$ are some appropriately chosen
set of homogeneous polynomials of degree $k$ in the $v_\ell$.
It then follows that $G_{\alpha, k}$ is a function of modular weight
$k-1$, and therefore can be written in terms of $E_4$ and $E_6$
alone.

The final ingredient is to invert $\phi_i(v_j,\tau)$,  to obtain
$v_j(\phi_i, \tau)$, and consider its modular properties.
The fact that such an inversion is possible follows from
the implicit fuction theorem and the fact that
at $m=0$ one has $\phi_j = \omega_1 v_j$, where $\omega_1$
is one of the periods of the base torus.  We now need to show
that $\phi(v_j,\tau)$ and hence $v_j(\phi_i, \tau)$ can be
written in terms of $E_2,E_4$ and $E_6$.

By definition, $\partial_{u_k}  \phi_j = \alpha_{jk}$
and  $\partial_{u_k}  \phi_{D,j} = \alpha_{D,jk}$, where
$(\alpha_{ij}, \alpha_{D,ij}) $ is the period matrix
of \foliate.  Hence:
\eqn\derivOm{{\partial \over \partial u_k}~ \big( \phi_{D,j} ~-~
\tau~\phi_j \big) ~=~ ( \alpha_{D,jk} ~-~ \tau~\alpha_{jk}) ~=~
\big( \widetilde \Omega_{ji} ~-~ \Omega_{ji} \big)~\alpha_{ik} \ .}
We have just seen that the left-hand-side has an expansion
in terms $E_4$ and $E_6$, and \hmatdefn\ along with the modular
properties of $h$ show that $( \widetilde \Omega_{ik} -
\Omega_{ik})$ has an expansion in terms of $E_2,E_4$ and $E_6$.
Therefore $\alpha_{ik}$ has such an expansion, which in turn
implies that $\phi_j(v_k, \tau)$ has such an expansion.

The conclusion is that when we substitute $v_j(\phi_i, \tau)$
into a series of the form \Lseries, one gets a series of the
form:
\eqn\newLseries{\phi_{D,i} - \tau \phi_i ~=~  \sum_{k = 0}^\infty ~
\sum_{\alpha} ~{\cal G}_{\alpha,k} (\tau) ~{m^{k+1}  \over
Q_{\alpha,k} (\phi_\ell)} \ ,}
where ${\cal G}_{\alpha,k}$ can be written in terms of $E_2$, 
$E_4$ and $E_6$,
and has a modular weight of $k-1$ (if one ignores the anomalous
modular behaviour of $E_2$).  Moreover, the $E_2$ dependence here
is inherited implicitly via the $v_j$.

The steps of the proof now go much as in section 2.
First, from the definition of $\phi_{D,i}$ and  $\phi_i$, one has
\eqn\indefints{\phi_i ~=~ \int^{u_k} ~ \alpha_{ij} ~du_j \ ,
\qquad  \phi_{D,i} - \tau\phi_i ~=~ \int^{u_k} ~
(\alpha_{D,ij} ~-~ \tau \alpha_{ij}) ~du_j \ .}
Since the $E_2$ dependence comes from the implicit dependence
in $u_k$, one has
\eqn\Etwopp{\eqalign{{\partial \big(\phi_{D,i} - \tau\phi_i \big)
\over  \partial E_2} ~=~ & (\alpha_{D,ij} ~-~ \tau \alpha_{ij})~
{\partial u_j  \over \partial E_2} ~=~ (\widetilde
\Omega_{ik} ~-~ \tau \delta_{ik})~
\alpha_{kj}~{\partial u_j  \over \partial E_2} \cr
{}~=~ & {\partial \big(\phi_{D,i} - \tau\phi_i \big)
\over  \partial \phi_k}~ \alpha_{kj}~{\partial u_j
\over \partial E_2}\ .}}

Differentiating the first equation in \indefints\ with
respect to $E_2$ yields:
\eqn\Etwosecond{0 ~=~  \alpha_{ij}~{\partial u_j
\over \partial E_2} ~+~ \int^{u_k} ~ \bigg({\partial \alpha_{ij}
\over \partial E_2}\bigg)_{u} ~du_j \ ,}
where the differentiation of $\alpha_{ij}$ is done with
respect to the explicit $E_2$ dependence, rather that
the implicit dependence via $u_j$. Here and below we adopt the
standard thermodynamic notation in which $(\del_x f)_y$ denotes
the derivative of $f$ with respect to $x$ holding $y$ constant.
{}From \derivOm\ and the fact that $\phi_D - \Omega \phi$ has
modular weight $-1$, it follows that $(\partial_{E_2}
(H~\alpha) )_u = 0$.  Moreover, inverting \hmatdefn\ one has
$(\partial_{E_2} H)_u = -{2 \pi i \over 12} H^2$, exactly as in
\Heq.   Combining these two facts one sees that:
\eqn\alpEtwo{H~\bigg({\partial \alpha \over \partial E_2}
\bigg)_u  ~=~  {2 \pi i \over 12}~H^2~\alpha \ .}
Using this, \Etwosecond\ and \indefints\  in \Etwopp, one finally
arrives at
\eqn\matrecur{\eqalign{{\partial \over  \partial E_2}\big(\phi_{D,i}
- \tau\phi_i \big) ~=~ &-{2 \pi i \over 24}~ {\partial
\big(\phi_{D,i} - \tau\phi_i \big) \over  \partial \phi_k} ~
\big(\phi_{D,k} - \tau\phi_k \big)  \cr ~=~ &-{2 \pi i \over 24}~
{\partial \over  \partial \phi_i} ~\big(\big(\phi_{D,k} - \tau\phi_k
\big) \big(\phi_{D,k} - \tau\phi_k \big) \big) \ .}}
from which the one also obtains a recursion relation like
\cFd\ for the pre-potential,
\eqn\cFdsun{
\dE\ctF=-\ {2\pi i\over24}\left(\partial_{\phi_i}\ctF\right)
\left(\partial_{\phi_i}\ctF\right).
}
 Note that in going to the
second equality of \matrecur, and hence to derive the
recursion relation for the the pre-potential, we crucially used the
integrability of the system:  that is, we used the symmetry of
${\partial  \phi_{D,i}  \over  \partial \phi_k}$ in $i$ and $k$.

\subsec{The $SU(3)$ instanton expansion}

With the recursion relation in \cFdsun, we can now start computing
the
instanton expansion.  The method is similar to that in section (2.2).
 The
perturbative contribution to the pre-potential is
\eqn\prepertsun{\eqalign{
\ctF_{pert}=-{1\over4\pi
i}\sum_{i<j}\biggl((\phi_{ij})^2\log(\phi_{ij})^2-\ha
\Bigl(&(\phi_{ij}+m)^2\log(\phi_{ij}+m)^2\cr
&+(\phi_{ij}-m)^2\log(\phi_{ij}-m)^2
\Bigr)\biggr),
}}
where $\phi_{ij}=\phi_i-\phi_j$.  As an expansion in $m^2$, we have
\eqn\prepertm{
\ctF_{pert}={1\over4\pi i}\sum_{i<j}\left((\log(\phi_{ij})^2+3)m^2+
{m^4\over6(\phi_{ij})^2}+{m^6\over30(\phi_{ij})^4}+
{m^8\over84(\phi_{ij})^6}+{\rm O}(m^8)\right)
}

The full pre-potential, as an expansion in $m$ has the form
\eqn\premexp{
\ctF={1\over 4\pi i}f_1(\tau,\phi_i)\ m^2-{1\over2\pi i}\sum_{n=2}
{f_n(\tau,\phi_i)\over2n-2}m^{2n}.
}
The recursion relation for the $f_n$ is then
\eqn\fnrec{
\dE f_n={n-1\over12}\sum_{m=1}^{n-1}{\left(\dpi f_m\right)
\left(\dpi f_{n-m}\right)\over(2m-2)(2n-2m-2)}.
}
Unlike the perturbative piece, which only has terms of the form
$(\phi_{ij})^{-2l}$, the instanton contributions will have terms that
combine
the different $\phi_{ij}$.  The recursion relation will explicitly
generate
such terms.  One then needs to adjust the $E_2$ independent pieces of
the $f_n$ so that the $q$ expansion has no leading order terms that
have
this mixed form.

The first four $f_n$ for the general $SU(n)$ theory are
\eqn\fnfour{\eqalign{
f_1&=\sum_{i<j}\log(\phi_{ij})^2\cr
f_2&={E_2\over6}\sum_{i<j}{1\over{\phi_{ij}}^2}\cr
f_3&=\left({{E_2}^2\over18}+{E_4\over90}\right)
\sum_{i<j}{1\over{\phi_{ij}}^4}-\left({{E_2}^2\over144}-
{E_4\over144}\right)
\sum_{i\ne j\ne k\ne i}{1\over{\phi_{ij}}^2{\phi_{ik}}^2}\cr
f_4&=\left({5{E_2}^3\over216}+{E_2E_4\over90}+{11E_6\over7560}\right)
\sum_{i<j}{1\over{\phi_{ij}}^6}-\left({{E_2}^3\over144}-
{E_2E_4\over240}- {E_6\over360}\right)\sum_{i\ne j\ne k\ne
i}{1\over{\phi_{ij}}^4{\phi_{ik}}^2}
\cr
&\qquad\qquad+
\left({{E_2}^3\over288}-{E_2E_4\over480}-{E_6\over720}\right)
\sum_{i<j<k}{1\over{\phi_{ij}}^2{\phi_{ik}}^2{\phi_{jk}}^2}\cr
&\qquad\qquad
+\left({{E_2}^3\over432}-{E_2E_4\over144}+{E_6\over216}\right)
\sum_{i\ne j\ne k\ne
l}{1\over{\phi_{ij}}^2{\phi_{ik}}^2{\phi_{il}}^2}
}}
Because of symmetry, the sums in \fnfour\ come with overall
integer factors.  Taking this into account, the $q$ expansions of these 
$f_n$ have coefficients
that are integer multiples of $2n-2$.  This leads us to conjecture
that the
expansion of $\del_\tau\ctF/m^2$ in $m^2/{M_i}^2$ and $q$, where the
$M_i$ are
the masses of the charged vector bosons, has only integer
coefficients.

The number of different types of terms in $f_m$ increases rapidly as
$m$ is increased, thus for relatively high rank gauge groups, the
expansions
become unwieldy.
There is another problem with the computation of higher rank gauge
groups
which involves the gaps.  For general $SU(n)$, when flowing from the
$N=4$ theory to the $N=2$ SYM, the scaling is $m^{2n}q=\Lambda^{2n}$,
with $\Lambda$ finite as $m\to\infty$.  Therefore, $f_{nm+1}$ has one
more gap
than $f_{nm}$ and so the number of gaps jumps by one for every $n$
terms
in the expansion in \premexp.  This presents a problem for $SU(7)$
and
higher.  For these groups, the $f_7$ term does not have a gap.  But
we
need more information than what is given by the perturbative
expansion in
order to find $f_7$.  This is because the $E_2$ independent modular
form that
is contained in $f_7$ has weight 12 and the space of such forms has
dimension
2.  Therefore, we have to know at least some of the one instanton
contribution
in order to
compute the coefficients in front of the ${E_4}^3$ and ${E_6}^2$
terms.
Likewise, when we get to $f_{6m+1}$ we will need to know some of the
contributions
of the first $m$ instanton terms in order to find the coefficients of
the
$m+1$ terms that span the modular forms of weight $12m$.

On the other hand, if the rank is low then the recursion relation is
very
useful and one can use {\it Mathematica} to generate the instanton
expansion
to high order.  Let us consider the case of $SU(3)$.  Instead of
writing the expansion in terms of $\phi_{ij}$, it is more convenient
to
use the invariants $U$ and $\Delta$, where
\eqn\Udelta{
U={\phi_{12}}^2+{\phi_{13}}^2+{\phi_{23}}^2\qquad\qquad
\Delta={\phi_{12}}^2{\phi_{13}}^2{\phi_{23}}^2.
}
We were able to compute the first 24 terms in the expansion of
\premexp,
which is enough information to completely determine the first eight
instantons.
The contribution of the first four are
$$\eqalign{
\cF_1&=\left( m^4{U^2\over2\Delta}-m^6{U\over\Delta}\right)q\cr
\cF_2&=\biggl(m^4 {3U^2\over2\Delta}+m^6\left({18U\over\Delta}-
{3U^4\over4\Delta^2}\right)+m^8\left({5U^6\over64\Delta^3}-
{3U^3\over4\Delta^2}-{27\over\Delta}\right)\cr
&\qquad\qquad
+m^{10}\left(-{5U^5\over16\Delta^3}+{9U^2\over\Delta^2}
\right)+m^{12}\left({5U^4\over16\Delta^3}-
{21U\over2\Delta^2}\right)\biggr)q^2\cr
\cF_3&=\biggl(m^4{2U^2\over\Delta}+m^6\left({116U\over\Delta}-
{4U^4\over\Delta^2}\right)+m^8\left({5U^6\over2\Delta^3}-
{96U^3\over\Delta^2}+{144\over\Delta}\right)\cr
&\qquad
+m^{10}\left(-{7U^8\over12\Delta^4}+{67U^5\over3\Delta^3}+
{26U^2\over\Delta^2}
\right)+m^{12}\left({3U^{10}\over64\Delta^5}-{U^7\over4\Delta^4}-
{161U^4\over2\Delta^3}+
{436U\over\Delta^2}\right)\cr
&\qquad
+m^{14}\left(-{9U^9\over32\Delta^5}+{207U^6\over16\Delta^4}-
{27U^3\over\Delta^3}-{396\over\Delta^2}\right)+
m^{16}\left({9U^8\over16\Delta^5}-{125U^5\over4\Delta^4}+
{278U^2\over\Delta^3}\right)\cr
&\qquad
+m^{18}\left(-{3U^7\over8\Delta^5}+{265U^4\over12\Delta^4}-
{682U\over3\Delta^3}\right)\biggr)q^3 
}$$
\vfill\eject
\eqn\suIIIF{\eqalign{
\cF_4&=\biggl(m^4{7U^2\over2\Delta}+m^6\left({332U\over\Delta}-
{45U^4\over4\Delta^2}\right)+m^8\left({1095U^6\over64\Delta^3}-
{2925U^3\over4\Delta^2}+{2025\over\Delta}\right)\cr
&\qquad
+m^{10}\left(-{777U^8\over64\Delta^4}+{5121U^5\over8\Delta^3}-
{4941U^2\over\Delta^2}\right)+m^{12}\biggl({2151U^{10}\over512
\Delta^5}-{7647U^7\over32\Delta^4}+{2118U^4\over\Delta^3}\cr
&\qquad+{4905U\over\Delta^2}\biggr)
+m^{14}\left(-{715U^{12}\over1024\Delta^6}+{2139U^9\over64\Delta^5}+
{585U^6\over4\Delta^4}-
{18297U^3\over2\Delta^3}+{3240\over\Delta^2}\right)\cr
&\qquad+
m^{16}\left({1469U^{14}\over32768\Delta^7}+{1157U^{11}\over
2048\Delta^6}-
{101781U^8\over512\Delta^5}+{108279U^5\over32\Delta^4}+
{8901U^2\over8\Delta^3}\right)\cr
&\qquad
+m^{18}\biggl(-{1469U^{14}\over4096\Delta^7}+{2587U^{10}
\over128\Delta^6}-{387U^7\over32\Delta^5}-{13323\over2\Delta^4}+
{14805U\over3\Delta^3}\biggr)\cr
&\qquad
+m^{20}\left({4407U^{12}\over4096\Delta^7}-{41145U^9\over512\Delta^6}
+{22203U^6\over16\Delta^5}-{1773U^3\over
2\Delta^4}-{9639\over\Delta^3}\right)
\cr&\qquad
+m^{22}\left(-{1469U^{11}\over1024\Delta^7}+{14937U^8\over128
\Delta^6}-
{2484U^5\over\Delta^5}+{10407U^2\over\Delta^4}\right)\cr
&\qquad
+m^{24}\left({1469U^{10}\over2048\Delta^7}-{3861U^7\over64\Delta^6}
+{21777U^4\over16\Delta^5}-{26229U\over4\Delta^4}\right)\biggr)q^4.
}}

We can easily obtain the $N=2$ $SU(3)$ SYM instanton expansion from
the mass deformed $N=4$ expansion by taking the limit $m\to\infty$
and $q\to 0$, and keeping finite $\Lambda^6=m^6q$.  The first eight
terms in this instanton expansion are
$$\eqalign{
\cF_1&=-\left({\Lambda^6\over\Delta}\right)U\cr
\cF_2&=-\left({\Lambda^{6}\over\Delta}\right)^2U\left({21\over2}-
{5\over16}\psi\right)\cr
\cF_3&=-\left({\Lambda^{6}\over\Delta}\right)^3U\left({682\over3}-
{265\over12}\psi+{3\over8}\psi^2\right)\cr
\cF_4&=-\left({\Lambda^{6}\over\Delta}\right)^4U\left({26229\over4}
-{217777\over16}\psi+{3861\over64}\psi^2-{1469\over2048}\psi^3\right)\cr
\cF_5&=-\left({\Lambda^{6}\over\Delta}\right)^5U
\left(220878-80709 \psi+
{53499\over8}\psi^2-{244347\over1280}\psi^3
+{4471\over2560}\psi^4\right)\cr
\cF_6&=-\left({\Lambda^{6}\over\Delta}\right)^6U\biggl(8201045-
{9420803\over2}\psi+{20093193\over32}\psi^2-
{48449263\over1536}\psi^3\cr
&\qquad\qquad\qquad\qquad
+{1358263\over2048}\psi^4-{40397\over8192}\psi^5\biggr)
}$$
\vfill\eject
\eqn\SUIIISYM{\eqalign{
\cF_7&=-\left({\Lambda^{6}\over\Delta}\right)^7U\biggl(
{2278827252\over7}-272724552\psi+{749523903\over14}\psi^2-
{3702963825\over896}\psi^3\cr
&\qquad\qquad\qquad\qquad
+{37821921\over256}\psi^4-{35234427\over14336}\psi^5+
{441325\over 28672}\psi^6\biggr) \cr
\cF_8&=-\left({\Lambda^{6}\over\Delta}\right)^8U
\biggl({108545170581\over8}-
{251496872289\over16}\psi+{274083715485\over64}\psi^2\cr
&\qquad\qquad\qquad\qquad-
{961670300877\over2048}\psi^3
+{205509716343\over8192}\psi^4-
{181914175245\over 262144}\psi^5\cr
&\qquad\qquad\qquad\qquad
+{10012215681\over1048576}\psi^6-{866589165\over
16777216}\psi^7\biggr),
}}
where $\psi=U^3/\Delta$.

\newsec{Recursion relations for other simply laced groups}

Even though the Donagi-Seiberg-Witten curve is not known, we can
still make a reasonable guess for a recursion relation in the
instanton
expansion for the $SO(2n)$ and $E_n$ groups.  Recall that the
recursion
relation for $SU(4)$ is
\eqn\suIVrec{
\dE\ctF=-{2\pi
i\over24}\sum_{i=1}^4\left(\del_{\phi_i}\ctF\right)
\left(\del_{\phi_i}\ctF\right).
}
The perturbative contribution to the pre-potential is
\eqn\suIVperp{\eqalign{
\ctF_{pert}&=-{1\over 8\pi i}\sum_{i<j}^4\Biggl(
(\phi_i-\phi_j)^2\log(\phi_i-\phi_j)^2-\half\Bigl(
(\phi_i-\phi_j+m)^2\log(\phi_i-\phi_j+m)^2\cr
&\qquad\qquad\qquad
+(\phi_i-\phi_j-m)^2\log(\phi_i-\phi_j-m)^2\Bigr)\Biggr).
}}
Since $SU(4)\simeq SO(6)$, we should be able to express the
pre-potential
and the recursion relation in terms of three $SO(6)$ variables $a_i$.
To this end let
\eqn\soVIvar{\eqalign{
a_1&=\half\left(\phi_1-\phi_2-\phi_3+\phi_4\right)\cr
a_2&=\half\left(-\phi_1+\phi_2-\phi_3+\phi_4\right)\cr
a_3&=\half\left(-\phi_1-\phi_2+\phi_3+\phi_4\right)\cr
a_4&=\half\left(\phi_1+\phi_2+\phi_3+\phi_4\right).
}}
Clearly the recursion relation becomes
\eqn\soVirec{
\dE\ctF=-{2\pi i\over24}\sum_{i=1}^3\del_{a_i}\ctF\del_{a_i}\ctF.
}
The perturbative piece has no $a_4$ dependence, and as a result
$\del_{a_4}\ctF=0$.  Hence, we can restrict the sum in \soVirec\ from
1 to 3.
The perturbative pre-potential in $SO(6)$ coordinates is
\eqn\soVIperp{\eqalign{
\ctF_{pert}&=-{1\over 8\pi i}\sum_{i<j}^3\Biggl(
(a_i-a_j)^2\log(a_i-a_j)^2+
(a_i+a_j)^2\log(a_i+a_j)^2\cr
&\qquad\qquad
-\half\biggl(
(a_i-a_j+m)^2\log(a_i-a_j+m)^2
+(a_i-a_j-m)^2\log(a_i-a_j-m)^2\cr
&\qquad\qquad+
(a_i+a_j+m)^2\log(a_i+a_j+m)^2
+(a_i+a_j-m)^2\log(a_i+a_j-m)^2\biggr)\Biggr).
}}

Given the form of the $SO(6)$ recursion relation, it does not take a
great
leap of faith to postulate that the recursion relation for general
$SO(2n)$
is
\eqn\soNrec{
\dE\ctF=-{2\pi i\over24}\sum_{i=1}^n\del_{a_i}\ctF\del_{a_i}\ctF.}
The perturbative pre-potential is the obvious generalization of
\soVIperp.  We can then construct an $m^2$ expansion as in \premexp\
for $\ctF$.  For general $SO(2n)$, the first three terms in the
expansion
are
\eqn\fnthree{\eqalign{
f_1&=\sum_{i<j}\log\left({\phi_i}^2-{\phi_j}^2\right)^2\cr
f_2&={E_2\over6}\sum_{i<j}\left({1\over(\phi_i-\phi_j)^2}+
{1\over(\phi_i+\phi_j)^2}\right)\cr
f_3&=\left({{E_2}^2\over18}+{E_4\over90}\right)
\sum_{i<j}\left({1\over(\phi_i-\phi_j)^4}+
{1\over(\phi_i+\phi_j)^2}\right)-\left({{E_2}^2\over144}-
{E_4\over144}\right) \times\cr
&\times
\sum_{i\ne j\ne k\ne
i}\left({1\over(\phi_i-\phi_j)^2(\phi_i-\phi_k)^2}+
{1\over(\phi_i-\phi_j)^2(\phi_i+\phi_k)^2}+
{1\over(\phi_i+\phi_j)^2(\phi_i+\phi_k)^2}\right)
}}

As for the $E_n$ gauge groups, $E_8$ has the same recursion relation
as $SO(16)$. However, the perturbative contribution to the
pre-potential is modified to include contributions from the $SO(16)$
spinor. This change in the perturbative pre-potential propagates
through the entire instanton expansion through the recursion
relation.

The $E_7$ recursion relation has a sum of an $SO(12)$ and an $SU(2)$
piece on the right hand side. The perturbative pre-potential has
contributions from the adjoints of each of these groups, plus a
contribution from the $(32,2)$ that fills out the $E_7$ adjoint.
$E_6$ has a sum of an $SO(10)$ and a $U(1)$ on the right hand side of
the recursion relation. The perturbative piece of the pre-potential
has the contribution of the $SO(10)$ adjoint, plus a contribution
from both $SO(10)$ spinors with opposite $U(1)$ charges.

\newsec{Discussion}

The recursion relation  in \cFdsun\ has a form that is reminiscent
of a renormalization group equation derived in \DP.  These authors
find that
\eqn\DPeq{
\del_\tau\cF={1\over4\pi i}\sum_{j=1}^n\oint_{a_j} t^2 d\xi \ ,}
where $t$ and $\xi$ are Riemann surface coordinates defined in
\foliate\ and \lswSUn, and the integrals are around the $a_j$ periods
of the surface \foliate. Expanding the right-hand side of \DPeq\ in
powers of $q$ leads to the instanton expansion. The authors
explicitly computed the entire two instanton contribution for any
$SU(n)$. The two instanton contribution has powers of $m$ up to
$m^{4n}$. So to some extent, the approach of \DP\ and our approach
are complementary; for a given instanton number they can find
contributions for arbitrarily high powers of $m$, while for a given
power of $m$ we know the contributions of arbitrarily high powers of
$q$.

There are many possible avenues for further development of the ideas
presented here. One open problem is how to go beyond the limit $h=6$,
where $h$ is the dual Coxeter number. Recall that if $h\le6$, then
one can use the perturbative expansion and the gaps to completely
determine, order by order, the $m$ expansion of $\cF$. For $h>6$ the
presence of a gap, and knowledge of all the lower order terms, is
insufficient to fix all the terms in the expansion. However, our
recursion relation does generate an overdetermined system of
equations for the coefficients in the instanton expansion,
particularly at higher orders. It is therefore conceivable that gaps
and consistency at higher orders in the instanton expansion could
resolve indeterminacy in the lower order terms. Thus, while
technically more complex, it is possible that our recursion relation
and the gaps could still determine the series for $h>6$.
It would also be valuable to see if one could combine our
recursion relation with \DPeq\ and get deeper insight into the
structure of the expansion.

It is hoped that the recursion relation in \soNrec, or the instanton
expansion derived from it,  might lead to the Donagi-Witten curve for
the $SO(2n)$ or $E_n$ theories.  This is not completely farfetched.
In
\MNW\ the instanton expansion for the $E_8$ string in the $\tau\to
i\infty$
limit was used to compute the corresponding Seiberg-Witten curve.  In
that
case, it was known that the expansion is made up of $E_8$ characters.
This put restrictions on the form of the curve, and by going to high
enough
instanton number, was enough to determine all coefficients in the
curve.
In the $SO(2n)$ case and $E_n$ case, we in principle know the
instantons,
so it is conceivable that we can go back and find the curve.

Another interesting question is whether the recursion relations found
here for the $N=4$ theories can be generalized to $N=2$ theories with
vanishing $\beta$-functions. For example, the $SU(2)$ gauge theory
with $N_f=4$, with $m_1 = m_2 = m; m_3 = m_4 =0$ has an identical
effective action to the one discussed here.  For general values
of the mass paremeters, if a
recursion relation exists, we would expect it to involve Jacobi
$\theta$-functions.   Another example is the non-critical string with
Wilson loops that break the $E_8$ global symmetry to a smaller
subgroup. This recursion relation, if it exists, should also involve
Jacobi $\theta$-functions.

Finally, there is the issue of the topological interpretation of
the pre-potential, and in particular, of the integers that appear in
the $\tau$ derivatives of $\cF_n$.   Unlike the $E_8$ non-critical
string, the expansions do not have the usual Gromov-Witten form.
However, there are two natural ways in which one
might hope to find such an explanation: either in terms of the
topology of instanton moduli space or in terms of topological
amplitudes of non-critical strings.

\goodbreak
\vskip2.cm\centerline{\bf Acknowledgements}
\noindent
This work is supported in part by funds provided by the DOE under
grant number DE-FG03-84ER-40168.

\listrefs

\vfill
\eject
\end